\documentclass[nofootinbib,prc,twocolumn,showpacs,showkeys,aps,superscriptaddress]{revtex4}
\usepackage{graphicx,color}
\usepackage{bm}

\newcommand{\req}[1]{(\ref{#1})}
\newcommand{\bel}[1]{\begin{equation}\label{#1}}
\newcommand{\belar}[1]{\begin{eqnarray}\label{#1}}

\def\epsk{\epsilon_k}
\def\mev{\;{\rm MeV}}

\begin{document}

\title{Fission of super-heavy nuclei: Fragment mass distributions
and their dependence on excitation energy}

\author{N. Carjan}
\affiliation{
Joint Institute for Nuclear Research, 141980 Dubna, Moscow Region, Russia}
\affiliation{
Horia Hulubei - National Institute for Nuclear Physics and Engineering,
P.O.Box MG-6, RO-76900, Bucharest, Romania}
\author{F.A. Ivanyuk}
\affiliation{
Institute for Nuclear Research, 03028 Kiev, Ukraine}
\author{Yu.Ts. Oganessian}
\affiliation{
Joint Institute for Nuclear Research, 141980 Dubna, Moscow Region, Russia}
\date{\today}


\begin{abstract}

The mass and total kinetic energy distributions of the fission fragments
in the fission of even-even isotopes of superheavy elements from Hs (Z=108)
to Og (Z=118) are estimated using a pre-scission point model. We restrict to
nuclei for which spontaneous fission has been experimentally observed.
The potential energy surfaces are
calculated with Strutinsky's shell correction procedure.
The parametrization of the nuclear shapes is based on Cassini ovals.
For the just before scission configuration we fix $\alpha$=0.98, what corresponds to $r_{neck}\approx 2$ fm, and take into account another four deformation parameters: $\alpha_1,\alpha_3,\alpha_4,\alpha_6$.
The fragment-mass distributions are estimated
supposing they are due to thermal fluctuations in the mass asymmetry degree
of freedom just before scission. The influence of the excitation energy of the
fissioning system on these distributions is studied.
The distributions of the total kinetic energy (TKE) of
the fragments are also calculated (in the point-charge approximation).
In Hs, Ds and Cn isotopes a transition from symmetric to asymmetric
fission is predicted with increasing neutron number N (at N$\approx$168).
Super-symmetric fission occurs at N$\approx$160. When the excitation energy
increases from 0 to 30 MeV, the peaks (one or two) of the mass distributions
become only slightly wider.
The first two moments of the TKE distributions are displayed as a function of
the mass number A of the fissioning nucleus. A slow decrease of the
average energy and a minimum of the width (at N$\approx$162) is found.

\end{abstract}

\maketitle

\section{Introduction}
Projects to measure fission fragment properties for the heaviest elements
($Z >$ 106) are underway at several heavy-ion facilities around the world.
The SHE Factory of JINR-Dubna \cite{SND}, which will produce its first intense
beam in 2019, the HIAF+CUBE of ANU-Canberra \cite{DH} and the tandem
accelerator of the JAEA-Tokai \cite{KN} are just few examples.
In this context, theoretical calculations are very useful. On one side they
can improve the design of such experiments and on the other side they provide
predictions to compare with data allowing to test various theoretical
assumptions.

The SHE are synthesized in heavy-ion induced complete fusion reactions at
Bass-barrier energies .
Depending on the target used, they are usually divided in "cold" and "hot"
fusion. In the first category enter closed shell $^{208}$Pb and $^{209}$Bi
targets leading to compound nuclei (CN) with $\approx$ 20 MeV excitation.
After evaporation of maximum two neutrons the system finds itself below
the fission barrier.
In the second category enter neutron-rich actinide (U - Cf) targets bombarded
with $^{48}Ca$ leading to compound nuclei with 35 to 45 MeV excitation.
Here the Coulomb repulsion of colliding nuclei, which is the main hindrance to
fusion, is weaker and therefore the CN cross section larger. Of course
higher excitation results in lower survival probability of CN but this is
compensated by relatively high fission barriers close to N=184 shell.
Nuclei with Z = 112 to 118 are produced with cross sections
several orders higher in "hot" than in "cold" fussion.

The experimental method commonly used is to separate in flight the residues,
formed after evaporation of neutrons and gamma rays, from the beam and implant
them into a detector. The spontaneous fission occurs at the end of an $\alpha$
decay chain or competing with $\alpha$ decay during the chain.
A promising alternative is to detect the primary fragments from the decay of
the excited composite system before its cooling down by evaporation
\cite{Kozulin}. In this way one can study the fission of SHE at moderate
excitation ($\approx$ 35 MeV). It is of course necessary to separate the
fusion-fission from the quasi-fission but the cross section is orders of
magnitude higher than for the evaporation residue (microbarn vs. picobarn).
From theoretical point of view it is therefore necessary to study both
spontaneous fission of SHE and their fission at moderate excitation energies.
This is the goal of the present work.

In the present study, an improved version of the scission point model
\cite{NPA15} that has confirmed its ability to describe the mass and
kinetic energy distributions of the fission fragments from the spontaneous
fission of the heaviest actinides for which
such distributions have been measured \cite{cluster16}, is used.
More recently, the same model was applied to long series of isotopes with
atomic number $Z$ from 110 to 126 in order to study general trends
\cite{NPA17}.
\begin{figure*}[!ht]
\includegraphics[width=0.65\textwidth]{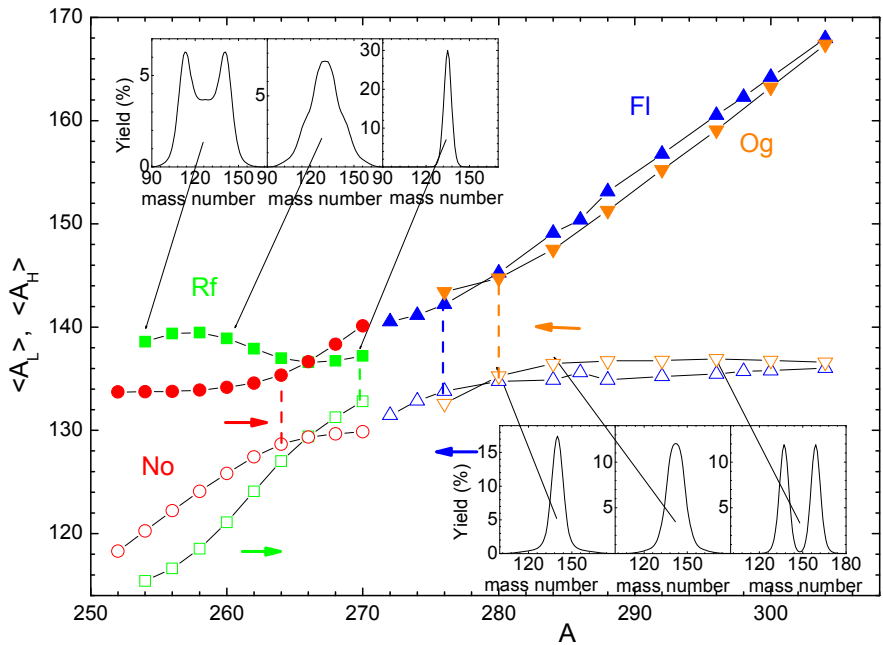}
\vspace{-3mm}
\caption{(Color online) Average masses of the light and heavy fragments
for isotopes of No, Rf, Fl and Og as a function of the mass number A of the
fissioning nucleus. The start points of the arrows mark the
transition from asymmetric to symmetric mass division and dashed verticals
mark the narrowest symmetric mass distribution in each series of isotopes.}
\label{fig1}
\end{figure*}
Thus, if one decreases the number of neutrons $N$ and includes the octupole
deformation $\alpha_3$, a transition from asymmetric to
symmetric mass division takes place in Fl, Lv, Og and $Z$=126 isotopes.
It is the mirror of the behaviour in Fm, No, Rf and Sg isotopes where the same
transition occurs with increasing $N$. In this way the fragment mass
systematics of the
SHE and of the heavy actinides join smoothly together as shown in Fig. 1.
It is a test of the reliability of the present approach.
It is interesting to notice that each series of isotopes has its narrowest
symmetric mass distribution with a full width at half maximum
 between 5 and 8 amu; hence extremely
small. To distinguish it from the regular symmetric fission,
we call this type of fission "super-symmetric".

The influence of the double magic $^{132}$Sn is clearly seen by the almost
constant mass ($\approx$ 136) of the heavy fragment in actinides and of the
light fragment in superheavies. The masses of the complementary fragment lie
on a straight line, simply reflecting the conservation of the total mass
number A.

However, the fission of many of these nuclei will never be observed and their
study is only academic.
Here we concentrate on nuclei with $Z >$ 106 for which spontaneous fission
has been
already detected although the statistics was not enough to build distributions.
As mentioned earlier, some of these nuclei will soon be remeasured in better
conditions. Our goal is to anticipate such experiments through a detailed
theoretical description of their fission fragments' properties.

Although the total kinetic energy of the fragments is also calculated, the
accent is put on their masses. Is the mass division expected to be symmetric
or asymmetric? How does the excitation energy of the fissioning nucleus
influences this mass division?
To answer such questions, we use
a pre-scission point model and therefore calculate the potential energy surface
(PES) of deformation, with Strutinsky's macroscopic-microscopic method
\cite{FH}, for a fissioning nucleus just-before its separation into two
fragments. These last mono-nuclear shapes are described by generalized Cassini
ovals with up to five deformation parameters $\{\alpha_i\}$. The corresponding
collective degrees of freedom (normal to the fission direction) are supposed
to be in statistical equilibrium. We therefore estimate the mass and TKE
distributions using Boltzman factors for the probability to populate the
points $\{\alpha_i\}$ on the PES.

As for the excitation energy dependence of PES, we a generalization of the
Strutinsky shell correction to finite temperature, suggested in Appendix A.
The generalization of the shell correction method is performed keeping the same
entropy (not temperature) in the original and averaged quantities, so that
the shell correction decreases monotonically, more or less exponentially with the
excitation energy.

The computational approach is explained in Sec. II. In Sec. III the predicted
fission fragment mass and TKE distributions are presented for the Hs, Ds, Cn,
Fl, Lv and Og isotopes for which spontaneous fission has been detected.
The effect of the excitation energy on the mass distributions is estimated.
A summary and conclusions can be found in Sec. IV.

\section{Computational details}
Our model is a "just-before scission" model \cite{NPA17} that uses
generalized Cassinian ovals
\begin{equation}\label{rx}
R(x)=R_0[1+\sum _n \alpha _n P_n(x)],
\end{equation}
to describe the ensemble of nuclear shapes involved. $R_0$ is the radius of
the spherical nucleus, $P_n(x)$ are Legendre polynomials
and $\alpha_n$ are the shape (deformation) parameters.
These shapes have in common a parameter $\alpha$ chosen such that at
$\alpha$=1.0 the neck radius is equal to zero irrespective of the values of
$\alpha_n$. For the just before scission configuration we fix $\alpha$=0.98 \cite{VP88}, and take into account another four deformation parameters $\alpha_1,\alpha_3,\alpha_4,\alpha_6$.

With the shape parametrization (\ref{rx}) we calculate the potential energy of
deformation using the microscopic - macroscopic approach \cite{FH}:
\begin{equation}\label{edef}
E_{def}=E_{def}^{LD} + \delta E,
\end{equation}
where $E_{def}^{LD}$ is the macroscopic liquid-drop energy including surface
and Coulomb energies
and $\delta E$ contains the microscopic shell and pairing corrections.

Each point ($\alpha_1,\alpha_3,\alpha_4,\alpha_6$) on the potential energy
surfaces has a certain probability to be realized.
Supposing statistical equilibrium for the collective degrees of freedom
normal to the fission direction \cite{NOR}, the distribution of these
probabilities is
\begin{equation}\label{boltz}
 P(\alpha_1,\alpha_3,\alpha_4,\alpha_6) \propto e^{-E_{def}(\alpha_1,\alpha_3,\alpha_4,\alpha_6)/T_{coll}} \,\,.
\end{equation}
Projecting
$P(\alpha_1,\alpha_3,\alpha_4,\alpha_6)$ on the fixed value
of mass asymmetry $\eta=(A_H-A_L)/A$ one obtains the fission fragment mass distribution $Y(\eta)$,
\begin{equation}\label{yield}
 Y(\eta)=\frac{\sum _{ijk} P(\alpha_1(\eta),\alpha_{3i}(\eta),\alpha_{4j}(\eta),\alpha_{6k}(\eta))\,}{\int d\eta \sum _{ijk}
P(\alpha_1(\eta),\alpha_{3i}(\eta),\alpha_{4j}(\eta),\alpha_{6k}(\eta))}\,.
\end{equation}
$T_{coll}$ is an unknown parameter that controls the overall width of the
distribution. In a sense it takes partially into account the dynamics.

In a similar way one obtains the fission fragment total kinetic energy (TKE)
distribution.
For each point ($\alpha_1,\alpha_3,\alpha_4,\alpha_6$) one calculates
the Coulomb interaction of the fragments
\begin{equation}\label{coul}
E_{coul}^{int} = e^2 Z_L  Z_H / D_{cm} = TKE
\end{equation}
Within a quasi-static approach one can not have access to the pre-scission
kinetic energy.
Including only the Coulomb repulsion energy into TKE, our estimates represent
lower limits.

The TKE distribution
\begin{equation}
 Y(TKE) = \sum _{ijkl} P(\alpha_{1i},\alpha_{3j},\alpha_{4k},\alpha_{6l})
\frac{e^{-\left({\frac{TKE_{ijkl}-TKE}{\Delta E}}\right)^2}}{\sqrt{\pi}\Delta E}\,,
\end{equation}
accounts for the finite energy resolution through the parameter $\Delta E$.

Here we will present only the first two moments calculated with the following
equations:
\begin{equation}
 <TKE>=\frac{\int TKE \,\, Y(TKE)\,\, dTKE}{\int  Y(TKE)\,\, dTKE}\,,
\end{equation}
and
\begin{equation}
\sigma^2_{TKE} =\frac{\int (TKE - <TKE>)^2\,\, Y(TKE)\, dTKE}{ \int  Y(TKE)\,\, dTKE}\,.
\end{equation}
\section{Predicted fission fragment mass and kinetic energy distributions}
The model described in the previous section is now applied to fragment
properties for superheavy nuclei for which spontaneous fission has been already
detected and could therefore be re-measured with better statistics in the near
future.
These are $^{264-278}$Hs, $^{268-280}$Ds, $^{276-286}$Cn,
$^{285-287}$Fl, $^{290-293}$Lv and $^{294}$Og \cite{OgUt,OgSoGur,Hess}.
\begin{figure}[!th]
\includegraphics[width=0.48\textwidth]{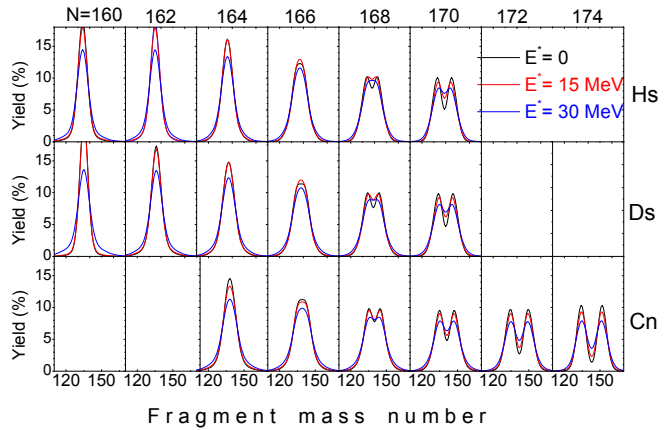}
\vspace{-5mm}
\caption{(Color online) The calculated fragment mass distributions
for isotopes of ${\rm Hs}$, ${\rm Ds}$ and ${\rm Cn}$
for which spontaneous fission has been detected. Three values of the
excitation energy E$^*$ have been considered. $T_{coll}$ = 2 MeV, E$_d$ =
40 MeV.}
\label{fig2}
\end{figure}

The calculated mass distributions corresponding to Hs, Ds and Cn isotopes are
presented in Fig. 2.
In all these three series, a transition from symmetric to
asymmetric fission is predicted with increasing neutron number N.
The transition point is N $\approx$ 168. The "super-symmetric" fission occurs
at N $\approx$ 160. The dependence on the excitation energy (calculated under assumption
of constant entropy) is also shown.
There is no noticeable difference between
E$^*$= 0 and 15 MeV and very small difference between E$^*$= 15 and 30 MeV. It is good
news since the features discussed above are not expected to be washed out if
these nuclei are produced with moderate excitation. So the mass distributions
in the SHE region is quite robust with respect to the excitation energy of
the fissioning nucleus.
The case of the heaviest nuclei ever produced, Fl, Lv and Og, is presented in
Fig. 3.
\begin{figure}[ht]
\includegraphics[width=0.48\textwidth]{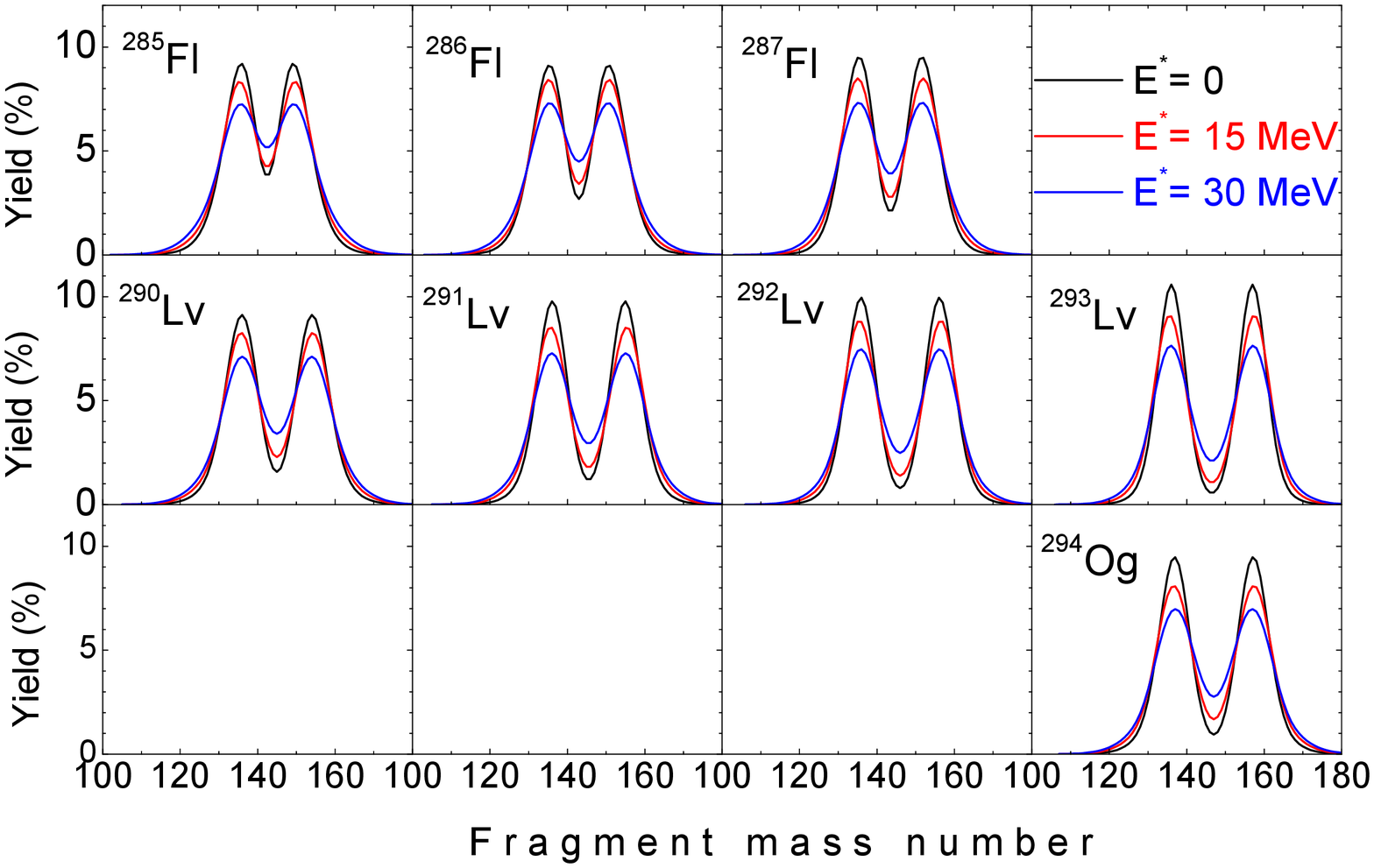}
\vspace{-5mm}
\caption{(Color online) The same as in Fig. 2 but for isotopes of ${\rm Fl}$,
${\rm Lv}$ and ${\rm Og}$.}
\label{fig3}
\end{figure}
All these nuclei are predicted to fission into fragments with unequal
masses. The dependence on the excitation energy is slightly stronger this time.
As expected, the mass-symmetric yield increases with E$^*$ but not enough to
overturn the mass-asymmetric character of the distributions.
There are already experimental indications that
$^{282,284,286}$Cn and $^{284,292}$Fl, at an excitation energy of about 30 MeV,
 may fission asymmetrically \cite{MI,EK1,EK2}.
These data are obtained by separating the fusion-fission and the quasi-fission
components. Since the error involved in this procedure is difficult to
assess, a measurement of spontaneous fission of the same nuclei would be
very welcome.

Fig. 4 shows the calculated mass distributions for a long series of
Ds isotopes. The heaviest three isotopes have not been detected but they are
candidates to be found together with Platinum in cosmic rays or in ores (as
eka-Pt)\cite{modane}. In agreement with the trends observed previously,
they are predicted to fission asymmetrically.

\begin{figure}[!tbh]
\includegraphics[width=0.48\textwidth]{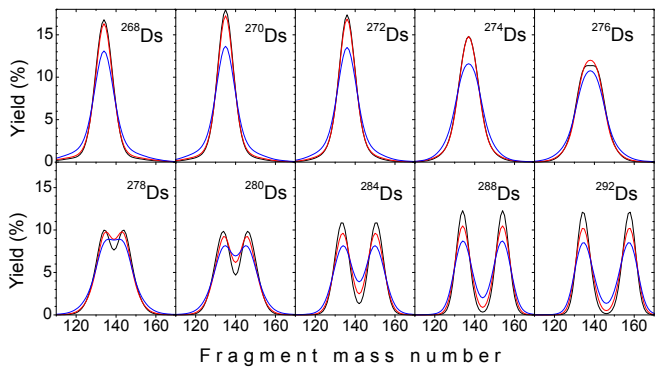}
\vspace{-5mm}
\caption{(Color online) The same as in Fig. 2 but for a larger series of
${\rm Ds}$ isotopes.}
\label{fig4}
\end{figure}
Let us now move to the total kinetic energy distributions. For all nuclei
studied here the shapes of TKE-distributions are identical (quasi-gaussian) and therefore a
similar presentation as for the mass distributions is not appropriate. Instead
we show the first two moments of these distributions as a function of the
mass A of the fissioning isotope in Figs 5 and 6 respectively.

The prescission
contribution to the average total kinetic energy is neglected meaning that the
values given are lower limits. For each series of isotopes there is a slight
decrease with A due to the increase of the radius (A$^{1/3}$) and a decrease
of the product $Z_L\times Z_H$. Concerning the width of the TKE distributions,
they exhibit a more complex variation with the mass of the isotope. For Hs and
Ds isotopes there is a minimum at N$\approx$162 close to the neutron number
where the mass distributions are also the narrowest. It is a sign of a strong
nuclear structure effect in extremely deformed nuclei (pre-scission shapes).
\begin{figure}[!tbh]
\includegraphics[width=0.45\textwidth]{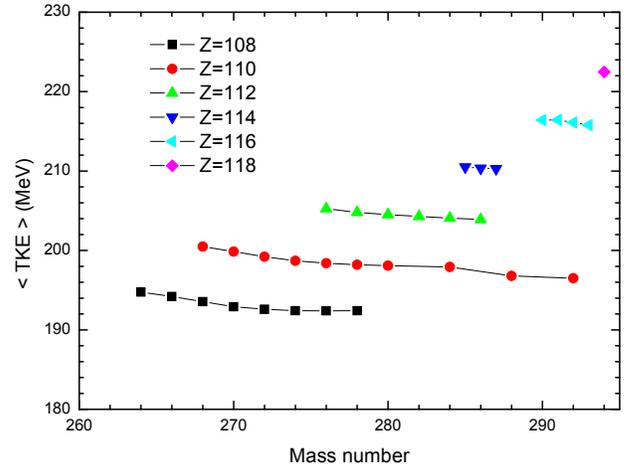}
\vspace{-3mm}
\caption{(Color online) The average total kinetic energy of the fragments
for the fission of isotopes of ${\rm Hs}$, ${\rm Ds}$, ${\rm Cn}$, ${\rm Fl}$,
${\rm Lv}$ and ${\rm Og}$
for which spontaneous fission has been detected.  $T_{coll}$ = 2 MeV,
$\Delta$E = 10 MeV.}
\label{fig5}
\end{figure}
\begin{figure}[h]
\includegraphics[width=0.45\textwidth]{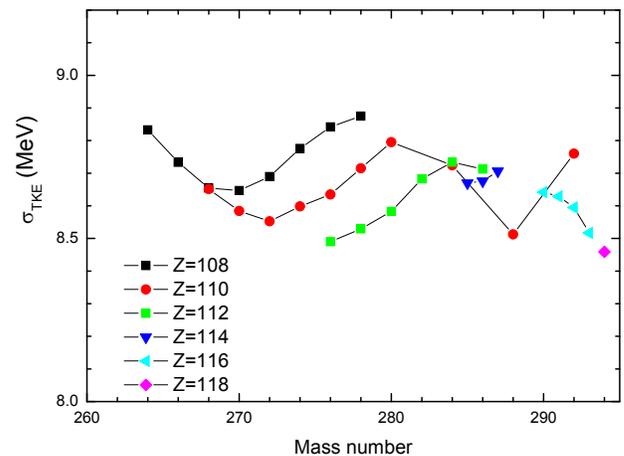}
\vspace{-3mm}
\caption{(Color online) Second moments of the total kinetic energy distributions
for the fission of isotopes of ${\rm Hs}$, ${\rm Ds}$, ${\rm Cn}$, ${\rm Fl}$,
${\rm Lv}$ and ${\rm Og}$
for which spontaneous fission has been detected.}
\label{fig6}
\end{figure}

\section{Summary and Conclusions}
\vspace{-3mm}
The mass and TKE distributions  of the fission fragments for the fission of
selected even-even isotopes of Hs, Ds, Cn, Fl, Lv and Og
are estimated using a pre-scission point model.
The influence of the excitation energy of the
fissioning system on these distributions is studied.
The underlying potential energy surfaces are calculated
with Strutinsky's shell correction procedure in a four dimensional
deformation space using Cassini ovals as basic shapes.
The results at finite excitation energies are obtained with a generalization
of this procedure keeping the same entropy in the original and averaged quantities,.

With increasing neutron number N, a transition from symmetric to asymmetric
fission is predicted at N$\approx$168.
Very narrow symmetric fission occurs at N$\approx$160. It is a signature
of a shell effect in extremely deformed nuclei.
There is no dramatic change in the mass distributions when the excitation
energy increases from 0 to 30 MeV; they are therefore quite robust.
With increasing the total mass A of the fissioning nucleus, a slow decrease
of the average total kinetic energy and a minimum (at N$\approx$162) of the
width of the TKE distribution is found.

We are therefore providing fission fragment distributions in a region that
will be soon explored experimentally.

\section*{Acknowledgements}
This work was partially done in the frame of the pro\-ject
PN-III-P4-ID-PCE-2016-0649 (contract nr. 194/2017).
One of us (F. I.) was supported in part by the project  "Fundamental
research in high energy physics and nuclear physics" of the National Academy
of Sciences of Ukraine.

\appendix
\section{The energy shell correction in excited nuclei}
The macroscopic-microscopic approach originally was developed for the nuclei at zero excitation energy (temperature). In \cite{strut67,strut68} the shell correction $\delta E_{shell}$ to the liquid drop energy of nucleus was defined as the difference between the sum $E(0)$ of
single-particle energies $\epsk$ of occupied states and some averaged quantity $\widetilde E$. In case of no pairing
\bel{deshell}
\delta E=E(0)-\widetilde E,
\end{equation}
where
\bel{eshell}
E(0)=\sum_{occ.}\epsk =\int_{-\infty}^{\epsilon_F} e g_S(e) de,\,\,g_S(e)\equiv 2\sum_k\delta(e-\epsk).
\end{equation}
The average part of energy is calculated by replacing in \req{eshell} the exact density of states $g_S(e)$ by the averaged quantity $\widetilde g(e)$,
\bel{etilde}
\widetilde E=\int_{-\infty}^{\tilde\mu} e \widetilde g(e)\,de\,.
\end{equation}

The generalization of Eqs. \req{deshell}-\req{etilde} to finite temperature is quite straightforward,
\bel{est}
E(T)=2\sum_k\epsk n_k^T, \, \rm{with}\, n_k^T=\frac{1}{1+e^{(\epsk-\mu)/T}}.
\end{equation}
The averaged energy $\widetilde E(T)$ is defined by replacing the sum over single-particle spectrum in \req{est}
by the integral with the smoothed density of states $\widetilde g(e)$,
\bel{etavr}
\widetilde E(T)=\int_{-\infty}^{\infty}e\,\widetilde g(e) n_e^Tde
\longrightarrow \delta E(T)=E(T)-\widetilde E(T),
\end{equation}
with $n_e^T\equiv 1/[1+e^{(e-\tilde\mu)/T}]$.
In the early works \cite{adeev72,adeev73,moretto72,aksel,bohrmo2,braque,civi82,civi83}  on the shell effects in excited nuclei and more recent publications \cite{ivahof,iva2018} it was shown that the shell correction to the {\it free} energy, $\delta F=\delta E- T\delta S$ ($\delta S$ is the shell correction to the entropy) decays more or less exponentially with the excitation energy. The temperature dependence of  shell correction to the energy $\delta E$ is more complicated. Up to temperature of approximately 0.5 $\mev$ the shell correction increases (in absolute value) with growing temperature and decays for larger temperatures. Such dependence is somewhat strange. Up to now there is no experimental information that the shell effects in excited nuclei may become larger compared with cold nuclei.

Looking at Fig. 2 in \cite{iva2018} it is clear that 
the rise (in absolute value) with temperature of the shell correction $\delta E$ to the energy comes from the shell correction to the entropy $\delta S(T)\equiv S(T)-\widetilde S(T)$.

 The shell correction to the entropy appears because in \cite{iva2018} the excited nucleus was considered as a canonical ensemble at some fixed temperature. I.e., the entropy $S$ and the average component $\widetilde S$ were calculated at the same temperature.

Meanwhile, it is clear that the nucleus is an isolated system. There is no thermostat which would keep the fixed temperature as the nucleus gets deformed. More meaningful would be to describe the excited nucleus by the microcanonical ensemble with energy and entropy as thermodynamical potentials.
I.e., the shell correction to the energy of excited nucleus should be calculated at fixed entropy,
not temperature. In other words, the energy $E$ and the energy $\widetilde E$ of the system with smoothed density distribution $\tilde g(e)$  should be calculated at the same entropy,
thus at {\it different} temperatures,
\bel{deltaes}
\delta E(S)=E(T(S))-\widetilde E(\widetilde T(S)),
\end{equation}
where the smoothed temperature $\widetilde T$ is defined by the requirement that the entropies calculated with exact $g_S(e)$ and averaged $\widetilde g(e)$ densities of states are the same,
\bel{tsmooth}
\widetilde S(\widetilde T)=S(T).
\end{equation}
\begin{figure}[ht]
\centering
\includegraphics[width=0.99\columnwidth]{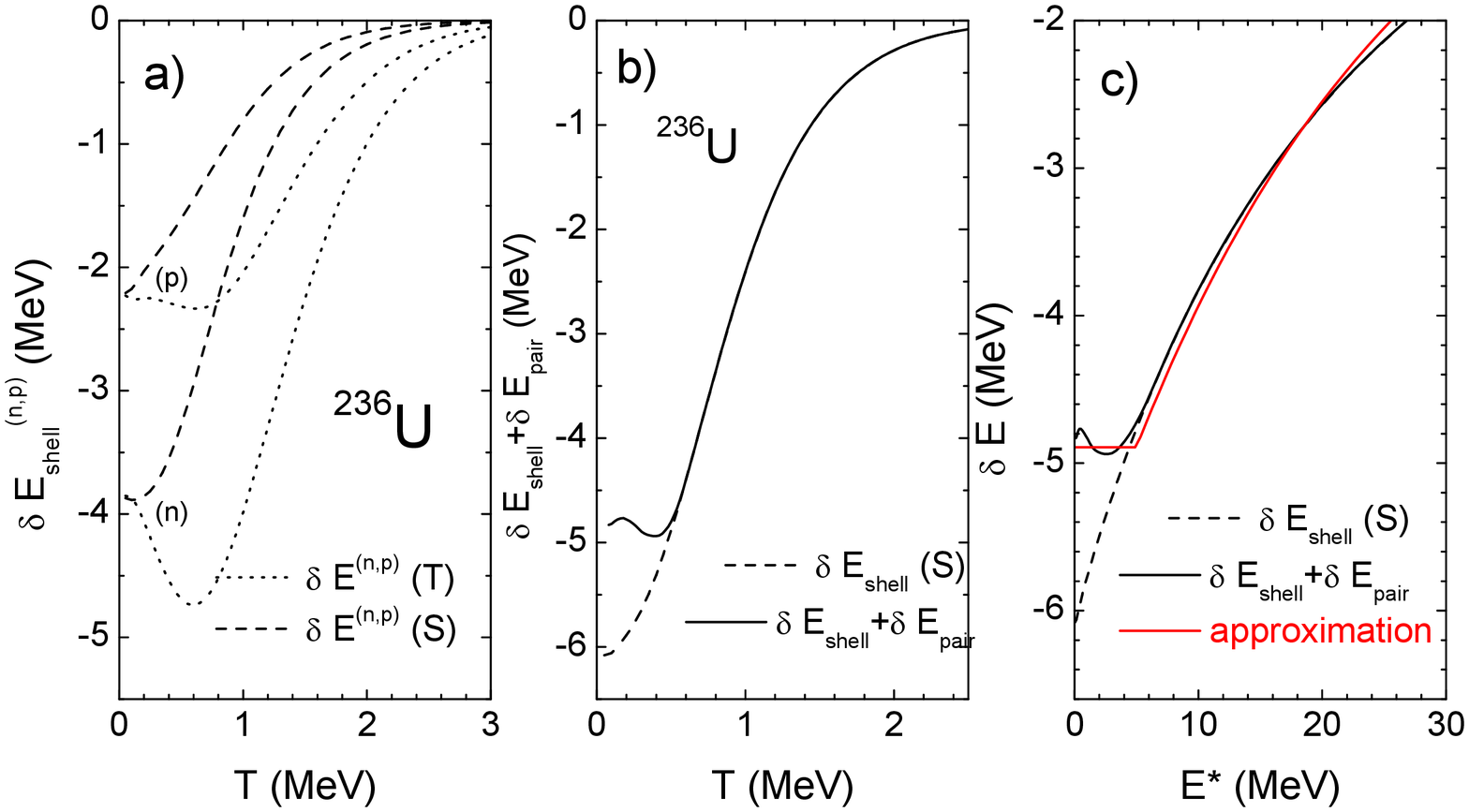}
\caption{(Color online) a) The shell corrections \protect\req{etavr} (dot) and \protect\req{deltaes} (dash) for protons (p) and neutrons (n) at the ground state of $^{236}U$ as function of temperature $T$;
b) The sum over protons and neutrons of the shell corrections \protect\req{deltaes} (dash) and \protect\req{deltae} (solid);
c) The shell corrections \protect\req{deltaes} (dash) and \protect\req{deltae} (solid) as function of the excitation energy. The red line is the approximation \protect\req{delefit}  with $E_d=23$ MeV.}
\label{A_fig1}
\end{figure}

The comparison of $\delta E(T)$ and $\delta E(S)$ is shown in Fig.\ref{A_fig1}a.
Unlike the $\delta E(T)$ the shell correction $\delta E(S)$ decreases monotonically with the growing temperature, what is quite reasonable.

The total shell correction (the sum over neutrons and protons) $\delta E_{shell}(S)$ is shown in Fig.\ref{A_fig1}b. In this calculation we include also the shell correction to the pairing energy,
\bel{deltae}
\delta E(S)\equiv \delta E_{shell}(S)+\delta E_{pair}(S).
\end{equation}
 The expressions for the pairing contributions to the averaged energy and entropy are given in \cite{iva2018}. As one could expect, the pairing correlations reduce the absolute value of the shell correction and up to the critical temperature the total shell correction $\delta E(S)$ does not depend much on the excitation energy. Above the critical temperature the total shell correction $\delta E(S)$ decreases approximately exponentially with the excitation energy.

Finally, in Fig.\ref{A_fig1}c we propose a simple approximation for the dependence of $\delta E(S)$ on the excitation energy $E^*$,
\belar{delefit}
\delta E(E^*)=\left\{
\begin{array}{rcl}
&\delta E(0),\,\,\quad\qquad\qquad\quad\\
&\rm{if}\,\vert\delta E_{shell}(0)e^{-E^*/E_d}\vert\ge \vert\delta E(0)\vert\,,\\
&\delta E_{shell}(0)e^{-E^*/E_d},\quad\\
&\rm{if}\,\vert\delta E_{shell}(0)e^{-E^*/E_d}\vert\leq \vert\delta E(0)\vert\,.
\end{array} \right.
\end{eqnarray}
In other words, up to the critical temperature, the shell correction is the same as at $E^*=0$ and decays exponentially for higher excitation energies.
In \req{delefit} the excitation energy is defines as the sum of excitation energies of neutrons and protons.

The use of approximation \req{delefit} makes it possible to avoid the very time consuming computations of the dependence of shell correction on the excitation energy. 
\begin{figure}[ht]
\centering
\includegraphics[width=0.99\columnwidth]{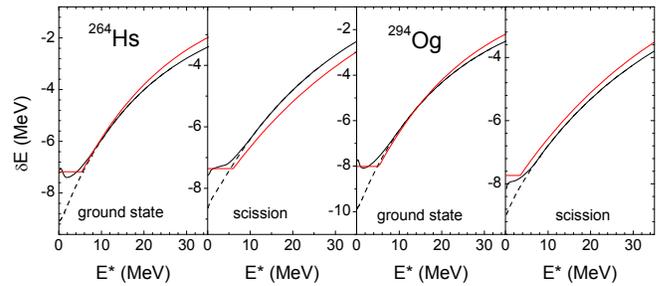}
\caption{(Color online) The calculated shell correction \protect\req{deltae} with (solid) and without account  of pairing correlations (dash) at the ground state and at scission ($\alpha=0.98, \alpha_n=0$) of $^{264}$Hs and $^{294}$Og and the approximation \protect\req{delefit} (red). The value of damping parameter $E_d=23$ MeV for the ground state and $E_d=40$ MeV for the scission. }
\label{A_fig2}
\end{figure}

In Fig.\ref{A_fig2} we compare the calculated dependence of the shell correction \req{deltae} on the excitation energy and the approximation \req{delefit} for the "light" and  "heavy" superheaavies $^{264}$Hs and $^{292}$Og at the ground state and at scission. One can see that the approximation \req{delefit} works reasonable well. Note, that in order to approximate the $\delta E(E^*)$ by \req{delefit} we had to use different values of the damping parameter, $E_d=23$ MeV at the  ground state and $E_d=40$ MeV at scission.

The more accurate approximation for the dependence of $\delta E$ on $E^*$ can be achieved with the two-parametric damping factor suggested in \cite{iva2018}. For the sake of simplicity we use in
calculations reported in the main text the exponential damping factor 
$\exp{(-E^*/E_d)}$ with the damping parameter $E_d=40$ MeV.

\end{document}